\documentclass[aps,prl,twocolumn,preprintnumbers,superscriptaddress,showpacs,floatfix]{revtex4-1}

\usepackage{epsfig}
\usepackage{bm}
\usepackage{amssymb}
\usepackage{amsmath}
\usepackage{color}
\usepackage{subfigure}
\usepackage{hyperref}
\usepackage{dcolumn}

\begin{document}

\title{Divergence of the Chapman-Enskog expansion in relativistic kinetic theory}
\date{\today}
\author{G.~S.~Denicol}
\affiliation{Instituto de F\'isica, Universidade Federal Fluminense, UFF, Niter\'oi, 24210-346, RJ, Brazil}
\author{J.~Noronha}
\affiliation{Instituto de F\'{\i}sica, Universidade de S\~{a}o Paulo, C.P. 66318,
05315-970 S\~{a}o Paulo, SP, Brazil}

\begin{abstract}
In this letter we show for the first time that the relativistic Chapman-Enskog series for a massless gas undergoing Bjorken expansion diverges. In order to fix this problem, we propose a novel type of expansion that includes non-perturbative contributions in the Knudsen number that are not considered in Chapman-Enskog theory. This approach is in good agreement with exact solutions of the Boltzmann equation for a wide range of values of Knudsen number and does not display the clear signs of divergence exhibited by the Chapman-Enskog series. 
\end{abstract}



\maketitle

\noindent \textsl{1. Introduction.} Relativistic formulations of fluid dynamics are widely used in physics with
applications in cosmology, high-energy nuclear physics, and astrophysics.
In the past decades, it has been the main theoretical framework used to describe
the time evolution of the quark-gluon plasma, a novel state of QCD matter produced in
ultrarelativistic heavy ion collisions performed at the Relativistic Heavy
Ion Collider (RHIC) and at the Large Hadron Collider (LHC). 
Fluid dynamical models of heavy ion collisions were decisive in the
discovery that the quark-gluon plasma behaves as a nearly perfect fluid with the smallest
kinematic viscosity known to exist in our universe (for a review, see \cite{Heinz:2013th}).

The onset of fluid dynamical behavior can be systematically studied from
first principles using the Boltzmann equation, as was first shown by
Hilbert, Chapman and Enskog \cite{hilbert,chapman-cowling} for the case of non-relativistic gases. In the Chapman-Enskog (CE) expansion \cite{chapman-cowling}, the single-particle distribution function is expanded
perturbatively in powers of the Knudsen number, $K_N$, a small parameter
given by the ratio between the mean free path and a characteristic
macroscopic length scale of the flow. This procedure leads to an
expansion of the distribution function in terms of gradients of the 
primary fluid-dynamical variables, namely temperature, chemical potential,
and velocity field. 

A distinct feature of the CE approach is that the 0th-order truncation of the
series leads to ideal fluid dynamics while its 1st-order
truncation leads to the well-known Navier-Stokes equations \cite{landau}, the most successful theory of fluid dynamics. However, second and third order truncations of the expansion in the non-relativistic regime lead to the so-called Burnett and super-Burnett equations \cite{burnett}, respectively, which were proven to be linearly unstable by Bobylev \cite{bobylev}. In the relativistic regime the situation is even more dire since
even the first order truncation of the series, i.e., the relativistic extension of
Navier-Stokes theory, is unstable due to its acausal nature \cite{hiscock}. Therefore, the applicability and convergence of the CE series in describing the dynamics of fluids is far from obvious.

As a matter of fact, even basic mathematical properties of this expansion, such as its convergence, are not fully understood. In the early 60's, Grad \cite{grad} proved that the 1st order truncation of the CE series is at least asymptotic to a solution of the Boltzmann equation, even if the whole series may diverge. Shortly after McLennan showed that under very specific conditions a convergent series can be found \cite{lennan}. A concrete example of a divergent CE series was found in \cite{dufty} for the case of uniform (stationary) shear flow for hard spheres using the relaxation time approximation \cite{bgk}. It was later shown that the inclusion of inelastic effects could render this type of series convergent \cite{santos}. So far, these analyzes have not been extended to consider expanding systems and, in the relativistic regime, such studies were never even performed.

In this letter we investigate the convergence of the relativistic CE series for a massless gas undergoing longitudinal Bjorken expansion \cite{bjorken}, a proxy for the dynamical evolution of the quark-gluon plasma formed in heavy ion collisions. In our approach, the Boltzmann equation is rewritten in terms of an infinite set of coupled ordinary differential equations for the moments of the distribution function \cite{Denicol:2012cn,Bazow:2015dha}, which are then expanded using the CE series. For the sake of simplicity, as in \cite{dufty}, we use the relaxation time approximation for the collision kernel of the Boltzmann equation. Under these circumstances, we demonstrate that the CE series diverges. We then propose a new type of series that includes both perturbative and non-perturbative contributions in the Knudsen number and leads to a reasonable approximation to solutions of the Boltzmann equation under Bjorken flow. This is the first study of the convergence of the CE series for an expanding gas.

\noindent
\textsl{2. Basics.} We consider a longitudinally expanding system of massless particles
undergoing Bjorken flow \cite{bjorken} in Minkowski spacetime, which is most conveniently described using the coordinates $x^\mu = (\tau, x,y,\eta)$ with $\tau=\sqrt{t^2-z^2}$ and $\eta = \tanh^{-1}(z/t)$ \cite{footnote1}. In these coordinates, the Bjorken expanding system is homogeneous and the local fluid velocity is $u^{\mu }=\left( 1,0,0,0\right)$, hence, all quantities depend solely on the time-like variable, $\tau$. However,
even though the fluid velocity is locally static, the system has
a nonzero expansion rate $\sim 1/\tau$ and, at
sufficiently early times, the gradients are large enough to drive
it far away from local thermodynamic equilibrium.

The state of this massless gas is specified by the single particle distribution 
$f_{\mathbf{k}}\equiv f\left(\tau,k_\eta,k_0\right) $ \cite{Denicol:2014xca,Denicol:2014tha}, where $k_{\mu }=(k_{0},\mathbf{k})$ is the 4-momentum of the particles and $k_{0}=\sqrt{k_{x}^{2}+k_{y}^{2}+k_{\eta
}^{2}/\tau ^{2}}$. The time evolution of $f_{\mathbf{%
k}}$ is determined by the Boltzmann equation in the following form
\begin{equation}
k_{0}\partial _{\tau }f_{\mathbf{k}}=\mathcal{C}[f].
\end{equation}%
In this work, we simplify the collision term using the relaxation time
approximation (RTA) \cite{bgk,AW},%
\begin{equation}
C\left[ f\right] =- u_{\mu }k^{\mu }\, \frac{f_\mathbf{k}-f_{\mathrm{eq}}}{%
\tau _{R}}=-k_{0}\frac{f_\mathbf{k}-f_{\mathrm{eq}}}{\tau _{R}},
\end{equation}%
where $\tau _{R}$ is the relaxation time and $f_{\mathrm{eq}}=\exp \left(
-k_{0}/T\right) $ is the local equilibrium distribution function with $T(\tau)$
being the temperature of the system. The temperature is related to the energy density $\varepsilon$ via the matching condition 
\begin{equation}
\varepsilon =\int \frac{d^{3}\mathbf{k}}{(2\pi )^{3}\tau }\text{ }k_0 f_\mathbf{k}\equiv \int \frac{d^{3}\mathbf{k}}{(2\pi )^{3}\tau }\text{ }k_0 f_{\mathrm{eq}}.
\end{equation} 
For a massless classical gas the condition above fixes $T$ in terms of the energy density $\varepsilon =3T^{4}/\pi ^{2}$. Also, in this system the thermodynamical pressure is given by $ P=\varepsilon/3$. 

Since the temperature enters in the expression for the collision term, the system is described by the coupled equations 
\begin{eqnarray}
\frac{\partial_{\tau}T}{T} +\frac{1}{3\tau}-\frac{\pi}{12\tau}=0,\label{conserv}\\
\partial _{\tau }f_{\mathbf{k}}=-\frac{f_\mathbf{k}-f_{\mathrm{eq}}}{\tau_{R}},\label{RTA}
\end{eqnarray}%
where \eqref{conserv} is the energy conservation equation (momentum conservation is trivially satisfied for the Bjorken flow profile). In this work, the relaxation time is chosen to be constant.
Also, $\pi\equiv \pi^\eta_\eta$ is a component of the shear stress tensor of the fluid, which is given by the following moment of $f_\mathbf{k}$
\begin{equation}
\pi_\eta^\eta=\int \frac{d^{3}\mathbf{k}}{(2\pi )^{3}\tau }k_0\left[\frac{1}{3}-\left( \frac{k_{\eta }}{k_{0}\tau }\right) ^{2 }\right]f_{%
\mathbf{k}}.
\end{equation}

Exact and semi-analytic solutions for the temperature and shear stress tensor in Bjorken flow have been found in Refs.\ \cite{Baym:1984np,Florkowski:2013lza,Florkowski:2013lya} while exact solutions in kinetic theory involving radial expansion can be found in \cite{Denicol:2014xca,Denicol:2014tha,Noronha:2015jia,Hatta:2015kia,Nopoush:2014qba}.  

\noindent
\textsl{3. Method of moments.} In this paper we use the relativistic version of the method of moments developed in
\cite{Denicol:2012cn} to solve Eqs.\ \eqref{conserv} and \eqref{RTA}. In this approach,
the state of the gas is described by moments of $f_\mathbf{k}$ and the
Boltzmann equation is formally replaced by an infinite set of coupled differential
equations of motion for these moments. Under Bjorken flow, the following moments encode all the microscopic information present in $f_\mathbf{k}$
\begin{equation}
\rho _{n,\ell }=\int \frac{d^{3}\mathbf{k}}{(2\pi )^{3}\tau }\left(
k^{0}\right) ^{n}\left( \frac{k_{\eta }}{k^{0}\tau }\right) ^{2\ell }f_{%
\mathbf{k}}.
\end{equation}%
The equation of motion for these moments is obtained by multiplying the
Boltzmann equation \eqref{RTA} by $\left( k^{0}\right) ^{n}\left(
k_{\eta }/k^{0}\tau \right) ^{2\ell }$ and integrating it with the measure $%
d^{3}\mathbf{k/}\left[ (2\pi )^{3}\tau \right] $. One then finds
\begin{equation}
\partial _{\tau }\rho _{n,\ell }+\frac{1+2\ell }{\tau }\rho _{n,\ell }+\frac{%
n-2\ell }{\tau }\rho _{n,\ell +1}=-\frac{1}{\tau _{R}}\left( \rho _{n,\ell
}-\rho _{n,\ell }^{\mathrm{eq}}\right) ,\label{equationrho}
\end{equation}%
where the equilibrium values for the moments can be expressed in terms of the temperature, 
\begin{equation}
\rho _{n,\ell }^{\mathrm{eq}}=\frac{\left( n+2\right) !}{2\ell +1}\frac{%
T^{n+3}}{2\pi ^{2}}.
\end{equation}%
Note that the matching condition used to define the temperature implies that 
$\varepsilon =\rho _{1,0}=\rho _{1,0}^{\mathrm{eq}}$. The initial
condition for the moments $\rho _{n,\ell }$ are calculated from the initial
condition for the distribution function and, using a moment expansion, $f_{\mathbf{k}}$ may be reconstructed at any given time \cite{Bazow:2016oky}.

In order to later implement the CE series, it is convenient to consider the dimensionless moments 
\begin{equation}
M_{n,\ell }\equiv \frac{\rho _{n,\ell }-\rho _{n,\ell }^{\mathrm{eq}}}{\rho
_{n,\ell }^{\mathrm{eq}}}.
\end{equation}%
whose equations of motion are derived from \eqref{equationrho} and read
\begin{gather}
\partial _{\tau }M_{n,\ell }+\frac{1}{\tau _{R}}M_{n,\ell }+\frac{6\ell -n}{%
3\tau }M_{n,\ell }-\frac{n+3}{12\tau }M_{1,1}\left(1+ M_{n,\ell}\right)  \notag \\
+\frac{1}{\tau }\frac{\left( n-2\ell \right) \left( 1+2\ell \right) }{2\ell
+3}M_{n,\ell +1}=-\frac{1}{\tau }\frac{%
4\ell \left( n+3\right) }{3\left( 2\ell +3\right) }.
\label{equacaoMnl}
\end{gather}%
This equation of motion makes the nonlinear nature of the RTA obvious since
the last term on the left-hand side depends on the product $%
M_{1,1}M_{n,\ell }$. Also, we note that the moment $M_{1,1}=-\pi /P$. 

What we obtained above is a hierarchy of equations of motion where, in order
to solve for $M_{n,\ell }$, we always need to know the next moment $M_{n,\ell
+1}$. We see that moments with different $n$ do not couple to each other,
implying that solving for the shear stress tensor requires knowledge of only
the moments $M_{1,\ell }$. We checked that the moment equations converge and agree with the exact solution found in Ref.\ \cite{Florkowski:2013lza}.

\noindent
\textsl{4. CE expansion.} In CE theory, one constructs a solution of the Boltzmann equation in terms of a power series in Knudsen number. For Bjorken flow, this
procedure leads to an expansion in powers of $K_N \sim \hat\tau^{-1}\equiv \tau _{R}/\tau $, where $%
1/\tau$ is the inverse macroscopic scale of the problem while $\tau _{R}$ corresponds to the
microscopic scale. Within the framework of the method of moments, this expansion scheme is implemented for the dimensionless moments $M_{n,\ell}$. Naturally, if this expansion diverges for $M_{n,\ell}$ it will also fail to converge for $f_\mathbf{k}$. 

The expansion for the moments has the following form
\begin{equation}
M_{n,\ell }=\sum_{p=0}^{\infty }\frac{\alpha _{p}^{\left( n,\ell \right) }}{%
\hat{\tau}^{p}},
\label{chupah}
\end{equation}%
where the dimensionless coefficients $\alpha _{p}^{\left( n,\ell \right) }$ are assumed to be independent of $\hat{\tau}$. These coefficients are obtained by substituting the expansion above into the equations of motion for the moments, $M_{n,\ell }$, Eq.\ \eqref{equacaoMnl}. This leads to  an algebraic and nonlinear equation for the coefficients $\alpha _{p}^{\left( n,\ell \right) }$. The general solutions are found by grouping the terms of the same power in $%
\hat{\tau}$ and solving for the coefficients order by order by iteration.

The lowest order coefficients are zero, $\alpha _{0}^{\left( n,\ell \right) }=0$, while the first order coefficients are
\begin{equation}
\alpha _{1}^{\left( n,\ell \right) }=-\frac{4\ell \left( n+3\right) }{%
3\left( 2\ell +3\right) }.
\end{equation}%
For $n,\ell=1$ one finds $\alpha_1^{(1,1)}=-16/15$, which leads to the following approximate expression for $\pi/P \approx 16/(15 \hat\tau)$. This is the same Navier-Stokes contribution commonly derived from kinetic theory \cite{Denicol:2010xn,Denicol:2011fa}.

The remaining higher order coefficients, $\forall m>1$, can be constructed
from the lower order coefficients using the formula%
\begin{eqnarray}
&&\alpha _{m+1}^{\left( n,\ell \right) }=-\frac{6\ell -n-3m}{3}\alpha
_{m}^{\left( n,\ell \right) }+\frac{n+3}{12}\alpha _{m}^{\left( 1,1\right) } \\ \nonumber  &-&
\frac{\left( n-2\ell \right) \left( 1+2\ell \right) }{2\ell +3}\alpha
_{m}^{\left( n,\ell +1\right) }+ \frac{n+3}{12}\sum_{p=0}^{m}\alpha
_{p}^{\left( 1,1\right) }\alpha _{m-p}^{\left( n,\ell \right) }.
\label{iteracao}
\end{eqnarray}%
This iterative procedure leads to an exact value for each coefficient. 

\begin{figure}[th]
\includegraphics[width=0.5\textwidth]{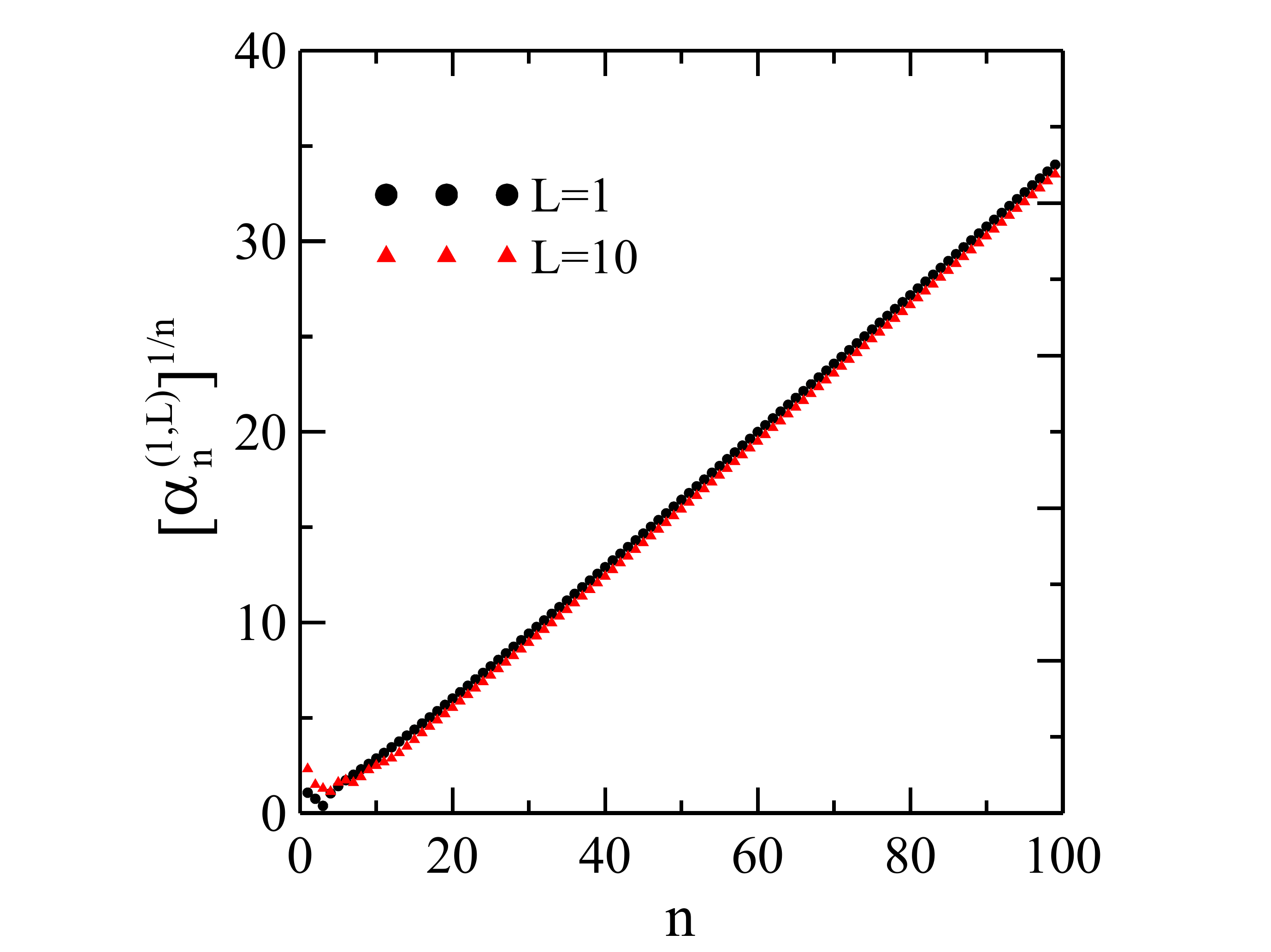}
\caption{(Color online)  $\left[ \alpha _{m}^{\left( 1,\ell \right) }\right] ^{1/m}$ as a function of $m$ for $L=1$ (black circles) and $L=10$ (red triangles).}
\label{chupa}
\end{figure}

In Fig.~(\ref{chupa}) we plot $\left[ \alpha _{m}^{\left( 1,\ell \right) }\right] ^{1/m}
$ for $\ell =1$ (which corresponds to the coefficients related to the
shear stress tensor) and $\ell =10$. For $m \gg 1$, we find
that this quantity grows linearly with $m$ indicating that $\alpha _{m}^{\left( 1,\ell \right) }\approx
m!$ independently of the value of $\ell$. We checked that the same holds for other values of $n$, which only affects the slope of the curve when $m$ is large. Therefore, in this case the CE expansion has zero radius of convergence.

In the nonrelativistic regime, Grad showed that the CE expansion, even if divergent, is asymptotic to a class of solutions of the Boltzmann equation \cite{grad}. In the relativistic regime, such proof has not been done even though it is generally expected that this is the case. Thus, if truncated appropriately, it may be possible for such series to provide a reasonable approximation to the distribution function in the small Knudsen number limit. On the other hand, the divergence of the series forbids its systematic improvement towards larger values of Knudsen number. This may be a problem when hydrodynamics is applied in extreme conditions such as in the theoretical description of heavy ion collisions where very large gradients exist at early times \cite{Niemi:2014wta,Noronha-Hostler:2015coa}, which also motivated the study of extended hydrodynamic descriptions \cite{Martinez:2010sc,Florkowski:2010cf,Bazow:2013ifa}.

\noindent
\textsl{5. Generalized CE expansion.} We now propose a way to improve the series in \eqref{chupah} by allowing the expansion coefficients to depend on $\hat\tau$
\begin{equation}
M_{n,\ell }\left( \hat{\tau}\right) =\sum_{p=0}^{\infty } \frac{\beta _{p}^{\left(
n,\ell \right) }\left( \hat{\tau}\right)}{\hat{\tau}^{p}}.
\end{equation}%
This will lead to a different class of solutions if $\beta _{p}^{\left(
n,\ell \right) }\left( \hat{\tau}\right)$ cannot be expanded in powers of $1/\hat\tau$. We shall see in the following that this is indeed the case due to the appearance of non-perturbative corrections $\sim \exp(-1/K_N)$. 

The main difference with respect to the calculation performed in the previous section is
that this new expansion leads to a set of coupled nonlinear differential equations for $\beta _{p}^{\left(
n,\ell \right) }(\hat\tau)$ instead of simple algebraic relations. In this case, the initial condition at $\hat\tau=\hat\tau_0$ for each coefficient must be extracted from $M_{n,\ell}(\hat\tau_0)$. Without loss of generality, we shall set $M_{n,\ell}(\hat\tau_0) =\beta_{0}^{\left(
n,\ell \right) }(\hat\tau_0)$ and take $\beta _{p>0}^{\left(
n,\ell \right) }(\hat\tau_0)=0$.

By collecting the terms of the same order in $\hat{\tau}$, we obtain the
differential equation satisfied by each $\beta _{m}^{\left( n,\ell \right) }(\hat\tau)
$. The lowest order coefficients satisfy a simple relaxation equation%
\begin{equation}
\partial _{\hat{\tau}}\beta _{0}^{\left( n,\ell \right) }+\beta
_{0}^{\left( n,\ell \right) }=0,
\end{equation}%
with solution%
\begin{eqnarray*}
\beta _{0}^{\left( n,\ell \right) }\left( \hat{\tau}\right)  &=&\beta
_{0}^{\left( n,\ell \right) }\left( \hat{\tau}_{0}\right) \exp \left(
-\Delta \hat{\tau}\right) , \\
\Delta \hat{\tau} &\equiv &\hat{\tau}-\hat{\tau}_{0}.
\end{eqnarray*}%
The remaining higher order coefficients, $\forall m>0$, satisfy the equation
of motion%
\begin{eqnarray*}
&&\partial _{\hat{\tau}}\beta _{m+1}^{\left( n,\ell \right) }+\beta
_{m+1}^{\left( n,\ell \right) } =-\frac{%
4\ell \left( n+3\right) }{3\left( 2\ell +3\right) }\delta_{m,0}\\ &-& \frac{\left( n-2\ell \right) \left(
1+2\ell \right) }{2\ell +3}\beta _{m}^{\left( n,\ell +1\right) }- \frac{%
6\ell -n-3m}{3}\beta _{m}^{\left( n,\ell \right) } \\ &+& \frac{n+3}{12}\beta _{m}^{\left( 1,1\right) }+\frac{n+3}{12}%
\sum_{p=0}^{m}\beta _{m-p}^{\left( 1,1\right) }\beta _{p}^{\left( n,\ell
\right) }.
\end{eqnarray*}%
Note that if $\partial _{\hat{\tau}}\beta _{m+1}^{\left( n,\ell \right) }=0
$ one finds $\beta \to \alpha$ and we recover the same result obtained with the traditional CE theory.

It is useful to first consider the simple example of a gas that is initially in local equilibrium where $\beta _{0}^{\left( n,\ell \right) }(\hat\tau)=0$. In this case, the first order equation is $\partial _{\hat{\tau}}\beta _{1}^{\left( n,\ell \right) }+\beta
_{1}^{\left( n,\ell \right) }=\alpha_{1}^{\left( n,\ell \right) }$, with $\alpha_{1}^{\left( n,\ell \right) }$ being the corresponding CE coefficient derived in the previous section. The solution is
\begin{equation}
\beta _{1}^{\left( n,\ell \right) }=\alpha_{1}^{\left( n,\ell \right)
}\left[ 1-\exp \left( -\Delta \hat{\tau}\right) \right].
\end{equation}%
Similarly,  one can find $\beta_2^{n,\ell}(\hat\tau)$ analytically
\begin{gather*}
\partial _{\hat{\tau}}\beta_{2}^{\left( n,\ell \right) }+\beta
_{2}^{\left( n,\ell \right) }=\alpha_{2}^{\left( n,\ell \right) }\left[
1-\exp \left( -\Delta \hat{\tau}\right) \right] , \\
\beta_{2}^{\left( n,\ell \right) }\left( \hat{\tau}\right) =\alpha%
_{2}^{\left( n,\ell \right) }\exp \left( -\Delta \hat{\tau}\right) \left[
\exp \left( \Delta \hat{\tau}\right) -\Delta \hat{\tau}-1\right] .
\end{gather*}%
A new feature of these generalized coefficients is that they contain contributions of the form $\exp(-j \,\hat\tau) \sim \left[\exp(-1/K_N)\right]^j$, with $j$ being a positive integer. Such terms display essential singularities when $K_N \to 0$ and, thus, can never be expanded in powers of the Knudsen number -- this is the mathematical reason behind the divergence of the usual CE expansion. We note that such non-perturbative terms were not assumed \textit{a priori}, but are naturally generated by the equations of motion derived for each expansion coefficient. Also, their presence is needed to describe the early time transient dynamics of the system.  We note that transport properties of the gas only enter the solution through $\hat\tau_0=\tau_0/\tau_R$. One can verify that the smaller $\hat\tau_0$ is, the larger $\pi/P$ becomes during the evolution of the system.

\begin{figure}[th]
\includegraphics[width=0.5\textwidth]{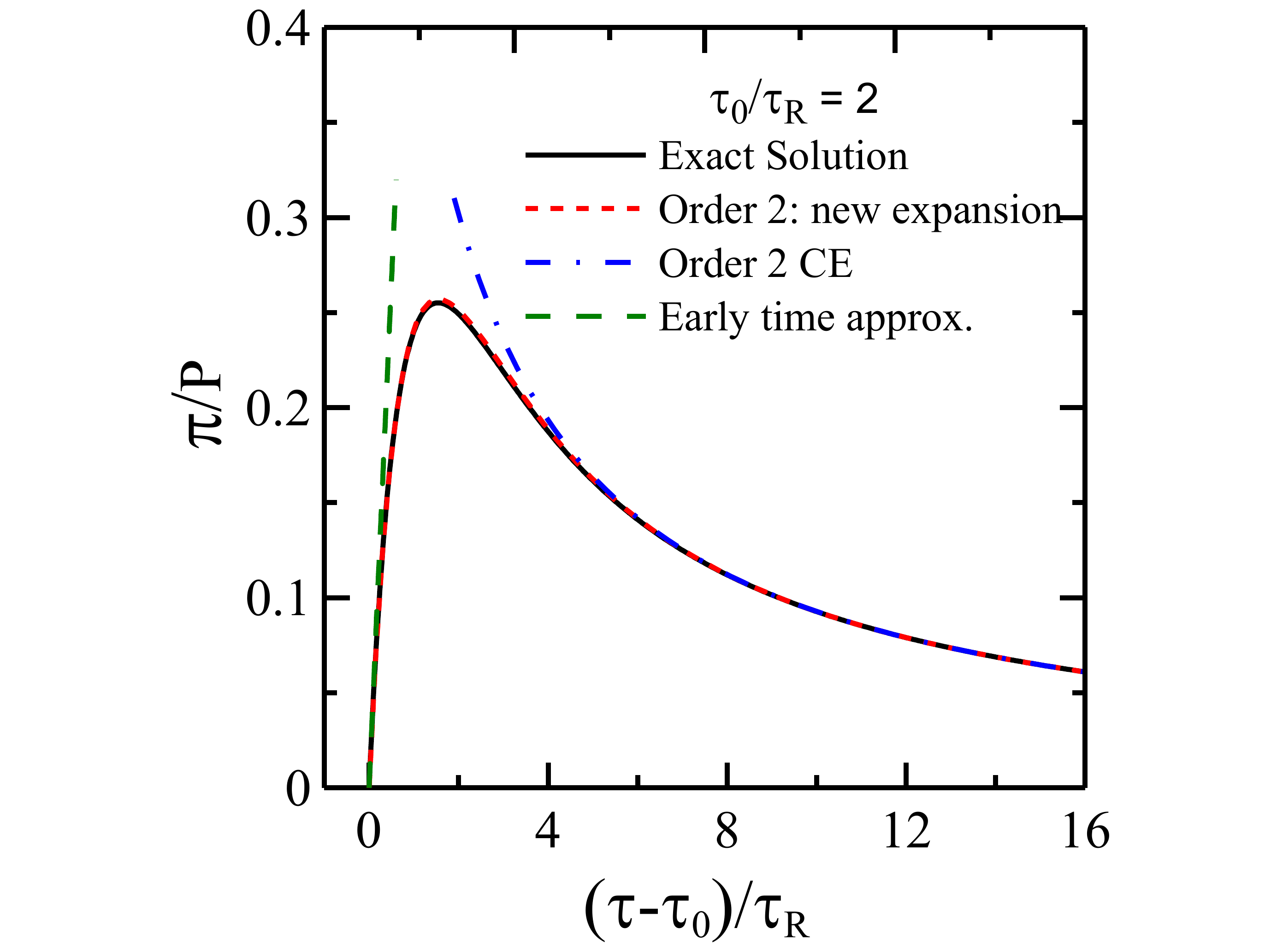}
\caption{(Color online) Comparison of several approximation schemes for $\pi/P$ as a function of $\Delta\hat\tau=(\tau-\tau_0)/\tau_R$ with the exact solution at $\hat\tau_0=2$.}
\label{chupareler}
\end{figure}

In Fig.~\ref{chupareler} we compare the second order solution, $\pi/P = -M_{1,1} \approx -\beta_1^{(1,1)}/\hat\tau-\beta_2^{(1,1)}/\hat\tau^2 $, for $\hat\tau_0 = 2$, with the exact solution obtained solving Eq.\ \eqref{equacaoMnl} and with the second order CE truncation. One can see that the generalized second order expression derived here is already in excellent agreement with the exact solution while the second order CE solution can only describe the late time dynamics. We note that as higher order corrections are included in the CE expansion the agreement with the exact solution worsens, even at late times, due to the divergence of the series.

We also include the $\Delta \hat\tau \ll 1$ limit of the solution for comparison. In this limit, one can show that $M_{n,\ell}$ is expressed as a series in terms of $\Delta \hat\tau$ instead of $1/\hat\tau$
\begin{equation*}
M_{n,\ell }\left( \hat\tau\right) =\alpha_{1}^{\left( n,\ell \right) }\frac{\Delta\hat\tau}{\hat{\tau}_{0}}+O\left( \Delta \hat{\tau%
}^{2}\right).
\end{equation*}%
We see that this expression is in good agreement with the exact solution at early times. This shows that at early times the system is better described by a power series in $1/K_N$ while at late times a series in powers of $K_N$ is more suitable. Our expansion seems to be able to capture both limits as it can be seen in Fig.~\ref{chupareler}.

\begin{figure}[th]
\includegraphics[width=0.5\textwidth]{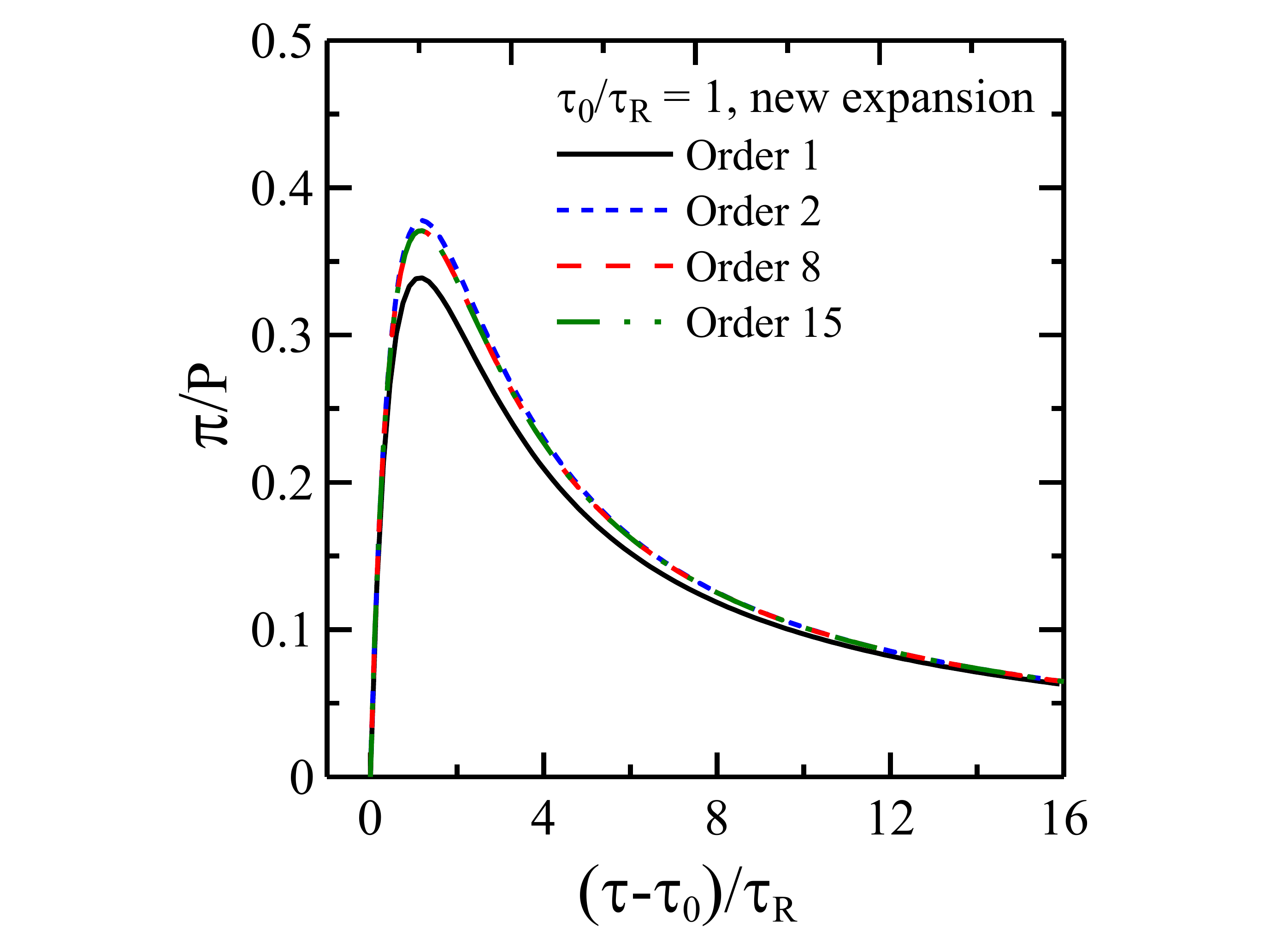}
\caption{(Color online) Test of the numerical convergence of $\pi/P$ for the generalized CE expansion.}
\label{chupareler2}
\end{figure}

Analytical expressions for the remaining higher order coefficients of the generalized expansion can also be found in this case, though the expressions are too long to be quoted here. In Fig.~\ref{chupareler2}, we illustrate the convergence of this generalized expansion by plotting the approximate solution up to 15th order, for $\hat\tau_0=1$. One can see that the new scheme appears to converge numerically already at second order, at least for $\hat\tau_0=1$.

We note that all these results were obtained for an initial state that is in local equilibrium. We have checked that basic properties, such as the apparent convergence of the approximate solution, remain valid for other initial conditions. 

\noindent
\textsl{6. Conclusions.} In this letter we showed that the relativistic CE series for a massless gas undergoing Bjorken expansion diverges. In order to fix this problem, we proposed a generalization of the CE expansion that also takes into account nonperturbative contributions in the Knudsen number. This formalism was shown to be in good agreement with exact solutions of the Boltzmann equation for a wide range of values of Knudsen number and, from a numerical point of view, appears to be convergent. This novel approach may contribute to our understanding of the emergence of fluid dynamical behavior at moderate to large values of the Knudsen number, as it occurs in heavy ion collisions.

The main limitation of this work has been the use of the RTA to simplify the collision term. Using the techniques developed in \cite{Bazow:2015dha,Bazow:2016oky}, it may be possible to perform a study similar to the one done here in the case where the full nonlinearities of the Boltzmann equations are taken into account. While we find it unlikely that such nonlinearities will affect the (lack of) convergence of the CE series in the Bjorken flow, they may affect the transient dynamics encoded in the new expansion derived here. 

The convergence of the gradient expansion has also been studied in the context of strongly coupled fluids defined within the gauge/gravity duality \cite{Maldacena:1997re} and, in the cases worked out in detail so far in holography \cite{Heller:2013fn,Buchel:2016cbj}, such an expansion was found to diverge. We note that the mathematical theory of resurgence has been employed to study this type of diverging series in recent works \cite{Heller:2015dha,Basar:2015ava,Buchel:2016cbj}. It would be interesting to apply these techniques also in kinetic theory and understand the relation between transseries expansions  and the generalized CE expansion proposed in this work.

\noindent
\textsl{7. Acknowledgements.} We thank M.~Martinez, U.~Heinz, and S.~Finazzo for discussions. JN thanks Conselho Nacional de Desenvolvimento Cient\'{\i}fico e Tecnol\'{o}gico (CNPq) and Funda\c c\~ao de Amparo \`{a} Pesquisa do Estado de S\~{a}o Paulo (FAPESP) for financial support.

\end{document}